# Single-shot deterministic complex amplitude imaging with a single-layer metalens

Liu Li[1,#], Shuai Wang[1,#], Feng Zhao[1], Yixin Zhang[1], Shun Wen[1], Huichao Chai[1], Yunhui Gao[1], Wenhui Wang[1], Liangcai Cao[1], Yuanmu Yang[1]*

[1]State Key Laboratory of Precision Measurement Technology and Instruments, Department of Precision Instrument, Tsinghua University, Beijing 100084, China
#These authors contributed equally.
*ymyang@tsinghua.edu.cn

**Conventional imaging systems can only capture light intensity. Meanwhile, the lost phase information may be critical for a variety of applications such as label-free microscopy and optical metrology. Existing phase retrieval techniques typically require a bulky setup, multi-frame measurements, or prior information of the target scene. Here, we proposed an extremely compact system for complex amplitude imaging, leveraging the extreme versatility of a single-layer metalens to generate spatially-multiplexed and polarization-phase-shifted point spread functions. Combining the metalens with a polarization camera, the system can simultaneously record four polarization shearing interference patterns along both in-plane directions, thus allowing the deterministic reconstruction of the complex amplitude light field in a single shot. Using an incoherent light-emitting diode as the illumination, we experimentally demonstrated speckle-noise-free complex amplitude imaging for both static and moving objects with tailored magnification ratio and field-of-view. The miniaturized and robust system may open the door for complex amplitude imaging in portable devices for point-of-care applications.**

Light wave contains both amplitude and phase information. However, due to its high oscillation frequency, approximately $10^{14}$ Hz[1], the phase of light cannot be directly detected by human eyes or photodetectors. Nonetheless, the simultaneous recording of the complex amplitude light field can provide comprehensive information towards various applications such as label-free microscopy in biomedicine[2,3], optical metrology[4,5], and adaptive optics[6].

To visualize phase information, Zernike first proposed the phase contrast technique[7] in 1934. Together with the later developed differential interference contrast[8] technique, they have been widely adopted in the observation of cells, while avoiding issues in fluorescence microscopy such as photobleaching and phototoxicity. Recently, there has also been a rapid development in quantitative phase imaging (QPI) techniques, which further allows the quantitative determination of morphological parameters, such as the thickness and refractive index, of the sample under test. One prevalent QPI technique is based on phase-shifting interferometry[9,10] or off-axis digital holography[11,12], in which the phase distribution can be retrieved from the interference pattern. However, despite their high accuracy and spatial resolution, these techniques remain prone to issues such as laser speckle noise and susceptibility to vibrations. Recent advancements based on Fourier ptychography[13,14], transport-of-intensity equation (TIE)[15], and lensless imaging[16-18] have made significant strides in addressing these limitations. Nonetheless, the phase reconstruction accuracy and speed of these improved techniques are often dependent on the number of measurement frames and may require specific sample priors. It remains a major challenge to construct a miniaturized system for the deterministic reconstruction of the complex amplitude light field in a single shot.

One potential strategy to address the abovementioned issue is to leverage the recent development in nanophotonics. Optical metasurface is an emerging class of diffractive optical elements consisting of nanoscale scatterers. Compared to conventional refractive or diffractive optical elements, it can be manufactured by a single-step lithography process and is extremely versatile in manipulating the vectorial light field[19-28]. Recently, metasurfaces[29-36] and thin-film optical devices[37-39] with nonlocal (angle-dependent) responses have been investigated for phase visualization and QPI. Most related to this study, a delicately aligned bilayer metasurface has been implemented to construct a quantitative phase gradient microscope[40], which can capture the one-dimensional (1D) phase gradient of the target object in a single shot. A single-layer metasurface has also been combined with a 4-$f$ imaging system to capture the 1D phase gradient based on polarization phase shift[41,42]. However, although the combination of iterative algorithms[43] and prior information of the target object can be used to reconstruct a two-dimensional (2D) phase image from a 1D phase gradient, the reconstruction is non-deterministic, with limited reconstruction fidelity and speed. There is still room to further reduce the overall volume of the imaging system.

In this study, we proposed and experimentally demonstrated an extremely compact imaging system using a single-layer metalens that allows the single-shot deterministic reconstruction of complex amplitude light field information. By judiciously designing the metalens with spatial- and polarization-multiplexed point spread functions (PSFs), the left- and right-handed circularly polarized (LCP and RCP) light shearing interference patterns can be recorded with a commercially available polarization camera. With a single measurement, four shearing interference patterns, each with a unique phase shift, were captured to generate phase gradient images along both orthogonal in-plane directions, with the verified smallest measurable phase gradient of 42 mrad/μm. Subsequently, the 2D phase gradient image can be used to deterministically reconstruct a 2D complex amplitude image with a relative phase accuracy of $0.0021\lambda$. Compared to conventional QPI techniques, our system is also immune to laser speckle noise with a light-emitting diode (LED) as the illumination source. We experimentally showcased that the system can be used for surface metrology and for dynamic complex amplitude imaging of living cells.

**Results**

**Working principle of the metalens-assisted single-shot complex amplitude imaging system.** The working principle of the metalens-assisted complex amplitude imaging system is schematically illustrated in Fig. 1a. The system was inspired by the polarization phase-shifting shearing technique, but has a greatly simplified form factor compared to conventional systems[44,45]. Using a single-layer metalens, a pair of shearing interference images of the target object, one along the $x$-axis, and another along the $y$-axis, can be simultaneously recorded on a polarization camera via spatial multiplexing. Each shearing interference pattern consists of LCP and RCP images with a shearing distance $\Delta s$ along the $x$- and $y$-directions, respectively.

The $x$- and $y$-shearing interference patterns are projected to the four polarization detection channels along 0°, 45°, 90°, and 135°, respectively, using the polarization camera (Fig. 1b). For each polarization channel, the recorded intensity of the shearing interference pattern can be calculated as (see Supplementary Note 1 for details),

$$I_m = I_{LCP} + I_{RCP} + 2\sqrt{I_{LCP}I_{RCP}}\cos\left[\Delta s \times \nabla_{x/y}\varphi + (m-1) \times \pi/2\right], \quad (1)$$

where $I_{LCP}$ and $I_{RCP}$ are the light intensity of the LCP and RCP images, respectively. $m = 1, 2, 3, 4$ corresponds to the polarization detection angle of 0°, 45°, 90°, and 135°, respectively. Subsequently, based on the four-step phase-shifting method, the phase gradients along the $x$- and $y$-direction $\nabla_{x/y}\varphi$ (Fig. 1c) can be deterministically calculated as,

$$\nabla_{x/y}\varphi = \frac{1}{\Delta s}\mathrm{atan}\left(\frac{I_{0°} - I_{90°}}{I_{45°} - I_{135°}}\right). \quad (2)$$

Finally, the 2D intensity image $I(x, y)$ can be extracted from the sum of the light intensities of two arbitrary orthogonal linear polarizations, while the 2D phase image $\varphi(x, y)$ can be reconstructed from the phase gradients using the higher-order finite-difference based least-squares integration (HFLI) method[46] (as shown in Fig. 1d, see Supplementary Note 2 for details). Note that the imaging system needs to be calibrated in prior with a collimated beam to eliminate the additional phase gradients caused by the non-planar wavefront of the incident beam and by possible system alignment errors (see Supplementary Note 3 for details). The system can use an LED as the illumination source to avoid laser speckle noise as long as the spatial coherence length is greater than the shearing distance $\Delta s$[47]. To avoid the random polarization phase difference between LCP and RCP light, the LED illumination is set to be linearly polarized.

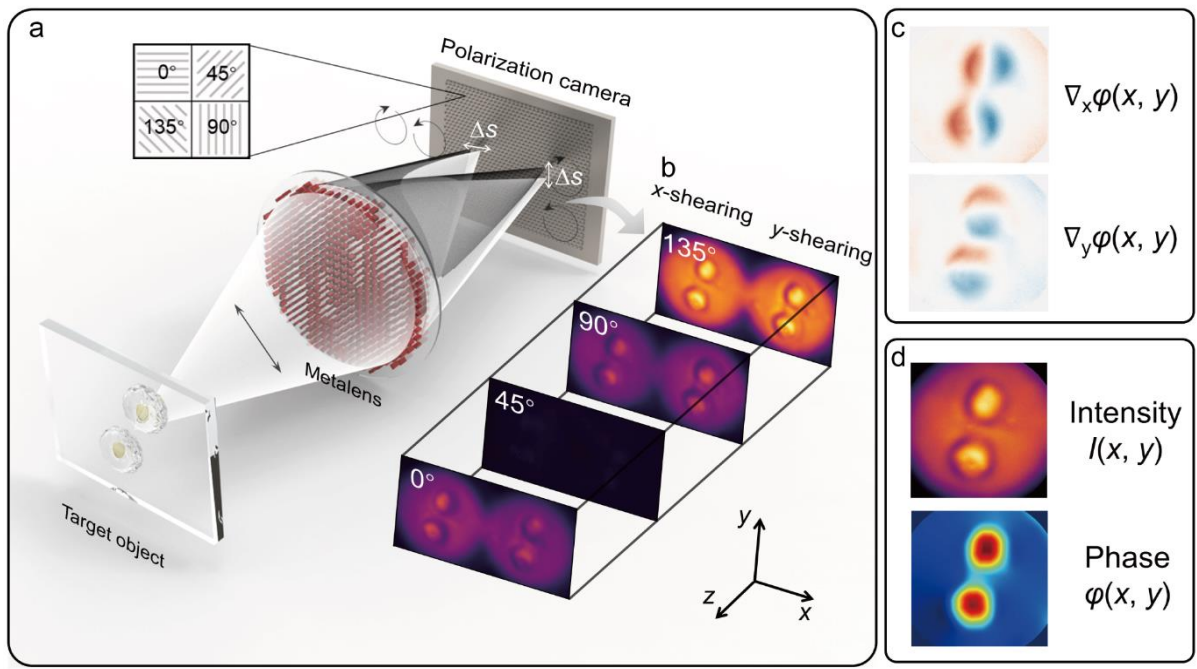

**Figure 1 | Working principle of the metalens-assisted single-shot complex amplitude imaging system. a,** Schematic of the metalens-assisted single-shot complex amplitude imaging system. **b,** Captured $x$-(left) and $y$-(right) shearing interference patterns with the polarization channel along 0°, 45°, 90°, and 135°, respectively. **c,** Calculated phase gradients along $x$- and $y$-direction, respectively. **d,** Reconstructed complex amplitude image from the 2D phase gradients.

**Metalens design and characterization.** The schematic of the spatial- and polarization-multiplexed metalens that shears and focuses orthogonally circularly polarized light is shown in Fig. 2a. With a plane wave incidence, the distance between the $x$- and $y$-shearing interference images is $d'$. With the target object placed at a finite distance, the actual image splitting distance $d$ can be tailored by varying the magnification ratio $M$ of the imaging system as $d = (1+M) d'$ and $\Delta s = (1+1/M) \Delta s'$, where $\Delta s'$ is the shearing distance with a plane wave incidence (see Supplementary Note 4 for details). For the phase reconstruction, the spatial sampling interval is directly proportional to $\Delta s$ and should ideally be minimized. Nonetheless, $\Delta s$ should be larger than twice the macro-pixel size of the polarization camera according to the Nyquist–Shannon sampling theorem[48]. For the metalens shown in Fig. 2, it has a diameter $D = 2$ mm and focal length $f = 1.5$ cm, resulting in a numerical aperture (NA) of 0.066. With collimated light illumination, the field-of-view (FOV) of the imaging system is identical to the aperture size of the metalens. The FOV can be further expanded with an increased lens diameter. For an object distance $d_o = 2.5$ cm and image distance $d_i = 3.75$ cm, the corresponding $M =$ 1.5, and the shearing distance $\Delta s$ = 25 μm. To avoid the overlap between the shearing interference images with the zero-order diffraction, the splitting distance $d'$ is set as 2 mm (see Supplementary Note 5 for details).

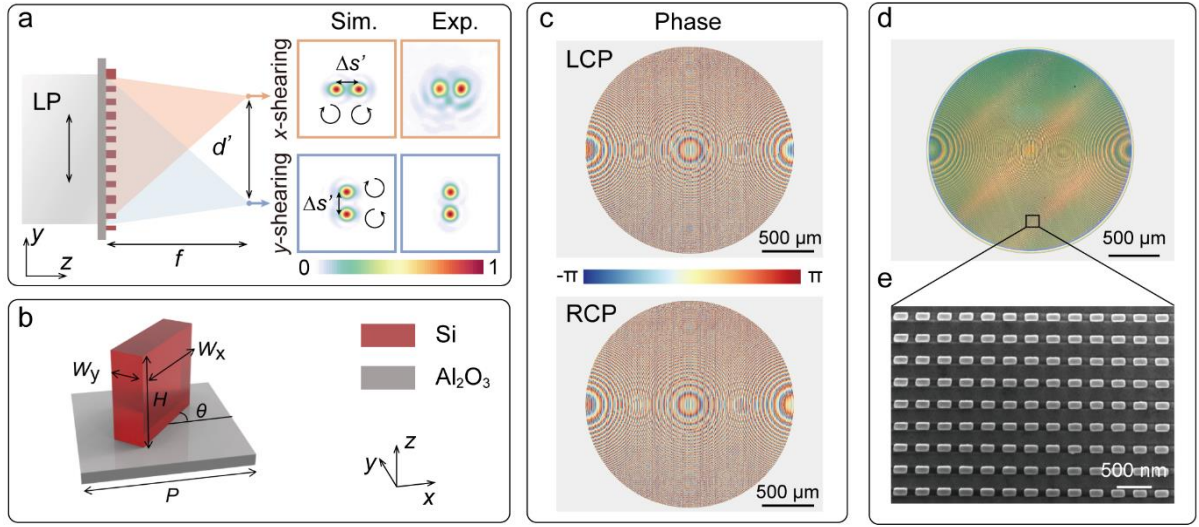

**Figure 2 | Metalens design and characterization. a**, Schematic of the metalens that splits and focuses the two pairs of shearing circularly polarized light on the photosensor, where $f$ is focal length, $d'$ is the distance between the $x$- and $y$-shearing interference images, and $\Delta s'$ is the shearing distance. The inset shows the simulated and measured PSFs along the $x$- and $y$-direction, respectively. **b**, The unit cell of the metalens is composed of rectangularly shaped silicon nanopillars on a sapphire substrate, with height $H$ = 600 nm and period $P$ = 350 nm. The rotation angle $\theta$ varies between 0 and $2\pi$. The width $W_x$ and length $W_y$ vary between 100 nm and 250 nm. **c**, Designed metalens phase profiles that can generate two pairs of shearing interference PSFs for LCP and RCP incident light, respectively. **d**–**e,** Optical microscopy image (**d**) and scanning electron microscopy image (**e**) of the fabricated metalens.

To generate such a shearing interference PSF at an operating wavelength of 800 nm, we designed a single-layer metalens consisting of an array of rectangularly shaped silicon nanopillars with varying geometry on a sapphire substrate, as shown in Fig. 2b. By combining the propagation phase of the rectangular nanopillars with the geometric phase generated through rotation, we can independently modulate the transmission phase of the LCP and RCP incident light over a full $2\pi$ range while maintaining near-unity transmittance. The phase profiles of the metalens for LCP and RCP incident light, which incorporates both focusing and polarization splitting terms, are depicted in Fig. 2c (see Supplementary Note 6 for details of the designed phase profiles). The designed metalens was fabricated using a standard electron-beam lithography and reactive-ion etching process (see Methods for details). The optical microscopy and scanning electron microscopy images of the fabricated metalens are shown in Fig. 2d, e, respectively. The measured PSFs for LCP and RCP incident are in close agreement with the simulation (inset of Fig. 2a). The metalens was measured to have a diffraction efficiency of 26.47% (see Methods and Supplementary Note 7 for details).

**Metalens-based compact complex amplitude imaging system.** The fabricated single-layer metalens was subsequently integrated with a polarization camera for complex amplitude imaging (see Methods for details), with an extremely small optical path volume of 0.85 × 0.71× 6.25 cm³, defined as the size of the photosensor × ( $d_o$ + $d_i$), as shown in Fig. 3a. Compared to conventional interference-based systems, the proposed system may be much less susceptible to environmental disturbance, such as mechanical shocks, air turbulence, and temperature fluctuations.

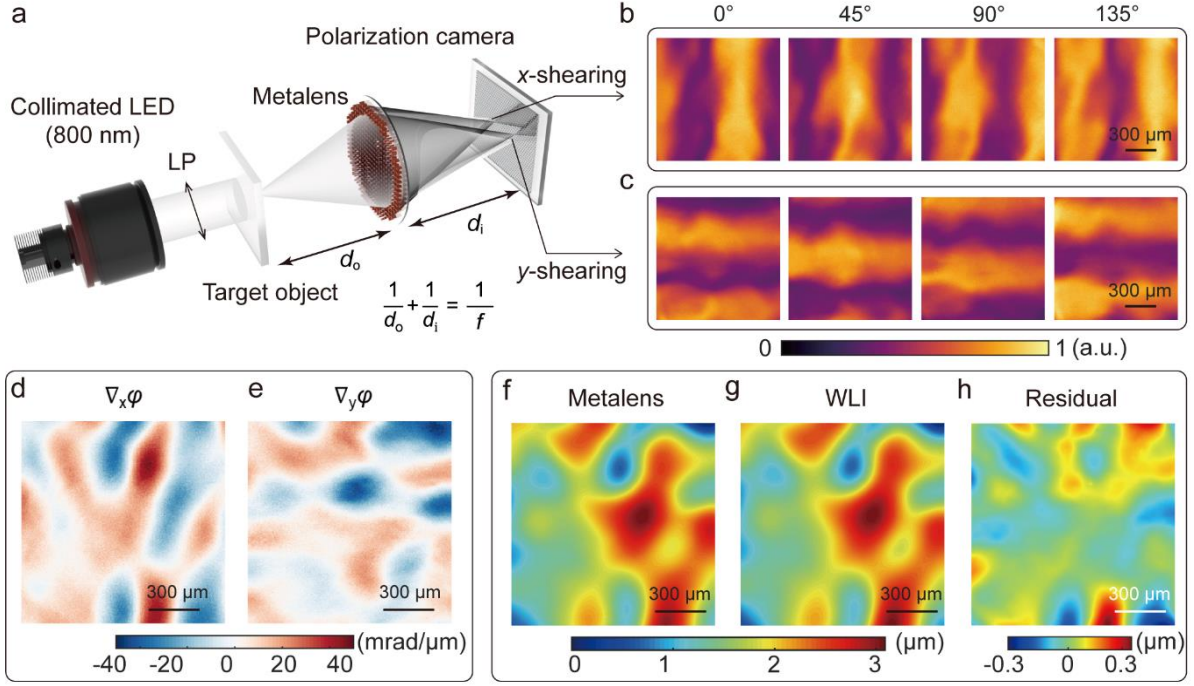

**Figure 3 | Surface metrology using the compact metalens-assisted single-shot complex amplitude imaging system. a**, Schematic of the compact complex amplitude imaging system. LP: Linear Polarizer. **b-c**, Captured *x*- (**b**) and *y*- (**c**) shearing interference patterns with the polarization channel along 0°, 45°, 90°, and 135°, respectively. **d-e**, Calculated phase gradients along the *x*- (**d**) and *y*-direction (**e**), respectively. **f-h**, Surface morphology of UV adhesive measured by the metalens-assisted system (**f**), by a commercial white light interferometer (WLI) (**g**), and the corresponding residual (**h**).

To experimentally validate the performance of the compact complex amplitude imaging system, we first measured the surface morphology of transparent drop-casted ultraviolet (UV) adhesives. The *x*- and *y*-shearing interferometric images in the four polarization channels were directly captured by the polarization camera, as shown in Fig. 3b-c. Subsequently, the phase gradients along both the *x*- and *y*-direction (Fig. 3d-e) and the 2D phase profile $\varphi(x, y)$ can be reconstructed. The resulting height distribution $z(x, y)$ of the UV adhesive sample is illustrated in Fig. 3f, which is derived as,

$$z(x, y) = \varphi(x, y) \frac{\lambda}{2\pi \Delta n}, \tag{3}$$

where $\lambda$ = 800 nm is the incident wavelength and $\Delta n$ is the refractive index difference between the UV adhesive sample and air.

To benchmark the system performance, we also measured the surface morphology of the identical UV adhesive sample using a commercial white light interferometer (WLI) (Nexview, Zygo), with the height distribution and residual shown in Fig. 3g,h, respectively. The relative average deviation (RAD) for the entire measurement area, defined as $\text{RAD} = \sum_{i=1}^{N} \left| \frac{(z_{i,\text{meta}} - \bar{z}_i)}{\bar{z}_i} \right| \frac{100\%}{N}$, is 3.79%, where $N$ is the number of spatial sampling points, and $\bar{z} = (z_{\text{meta}} + z_{\text{WLI}})/2$ is the average height measured by our system and by the WLI. While it took the WLI 15 s to measure the height distribution of the 1.2 × 1.2 mm² sample, it only took 10 ms for our system. Furthermore, we found the metalens-assisted imaging system could also be used to characterize the phase distribution of other phase objects, such as a metasurface carrying orbital angular momentum (see Supplementary Note 8 for details).

**Metalens-based complex amplitude microscopy.** The metalens can also be integrated with a standard microscope as a quick-release module for the observation of microscopic objects (see Methods for details), as shown in Fig. 4a. For microscopy applications, we designed a different metalens with a focal length of 2.5 mm, resulting in an NA of 0.37. With $d_o$ set as 3.9 mm, the corresponding $M = 1.75$ and $\Delta s = 1.57$ μm. After the secondary magnification using a 20× objective lens, the total system magnification ratio is 35.

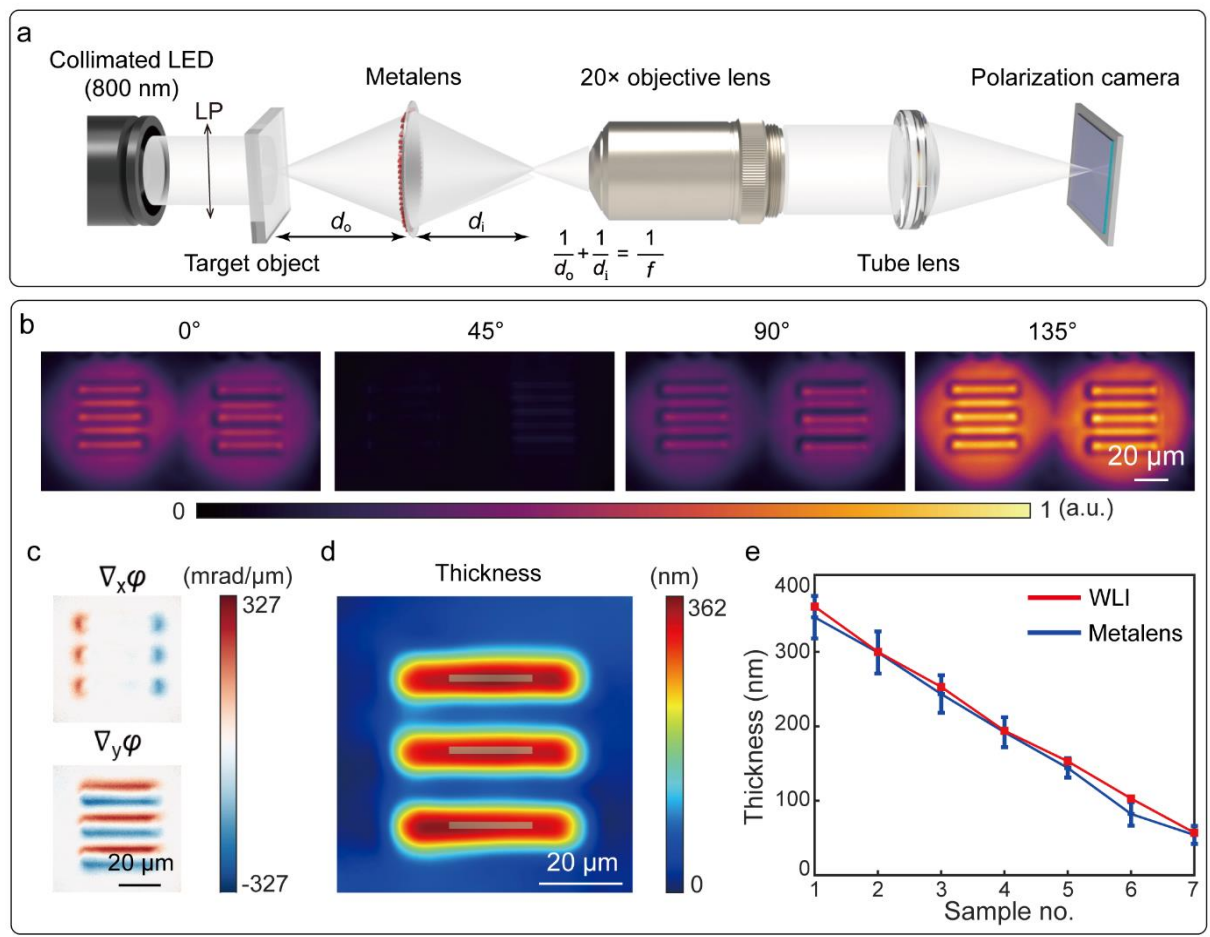

**Figure 4 | Characterization of the metalens-assisted complex amplitude microscopy system. a**, Schematic of the metalens-assisted complex amplitude microscopy system. LP: Linear Polarizer. **b**, Captured shearing interference images with the polarization channel along 0°, 45°, 90°, and 135°, respectively. **c**, Calculated phase gradients along the *x*- and *y*-direction, respectively. **d**, Reconstructed thickness of the 361-nm-thick phase resolution target (Group 6, Element 1). **e**, Comparison between the measured thickness of seven different phase targets by the proposed microscopy system and by a commercial WLI. Error bars represent standard deviations of the measured values.

To quantitatively characterize the phase reconstruction performance of the proposed microscopy system, we measured the phase distribution of a commercially available 1951 USAF phase resolution target (Quantitative Phase Microscopy Target, Benchmark Technologies). As shown in Fig. 4b, the four polarization shearing interference patterns of the 361-nm-thick resolution target (Group 6, Element 1) were captured in a single shot, allowing the subsequent determination of the phase gradients (Fig. 4c) and target thickness (Fig. 4d). Such a process was repeated for seven phase resolution targets with different thicknesses (see Supplementary Note 9 for details). The measured thickness, averaged over the gray area as depicted in Fig. 4d, closely matches the results obtained by a WLI, as shown in Fig. 4e. The verified smallest detectable phase gradient is 42 mrad/μm, corresponding to a phase resolution target at a height of 57 nm (see Supplementary Note 9 for details). In addition, the measured average spatial and temporal noise levels are 25.93 mrad/μm and 2.88 mrad/μm, respectively. With the phase reconstruction algorithm filtering the high-frequency spatial noise[46], the system has a high relative phase accuracy of 0.0021λ, defined as the root-mean-square error of the reconstructed reference plane wave[49] (see Supplementary Note 10 for details). The maximum detectable phase gradient $\nabla\varphi_{max}$ of the microscopy system is determined by the NA of the metalens and the shearing distance. In the current implementation, $\nabla\varphi_{max} = 2\pi/\lambda \cdot NA = 1.45$ rad/μm.

The complex amplitude microscopy system's spatial resolution for intensity and phase imaging was experimentally measured to be 4.38 μm and 5.52 μm, respectively (see Supplementary Note 11 for details), which is mainly limited by the sheared PSF and the diffraction limit. To further improve the spatial resolution of the imaging system, one could design a metalens with a higher NA or use advanced algorithms to further optimize the metalens topology[50]. In the microscopy system, the FOV is 85 × 85 μm$^2$, mainly limited by the size of the photosensor, the spatial multiplexing scheme, and the overlap between the zero-order diffraction and the shearing interference patterns. The FOV limitation caused by spatial multiplexing may be partially addressed by radial shearing interferometry, despite that such a scheme may result in uncertainty for the reconstruction of phase objects of particular topology[51]. The crosstalk between the zero-order diffraction and shearing interference patterns was avoided by off-axis imaging (see Supplementary Note 5 for details).

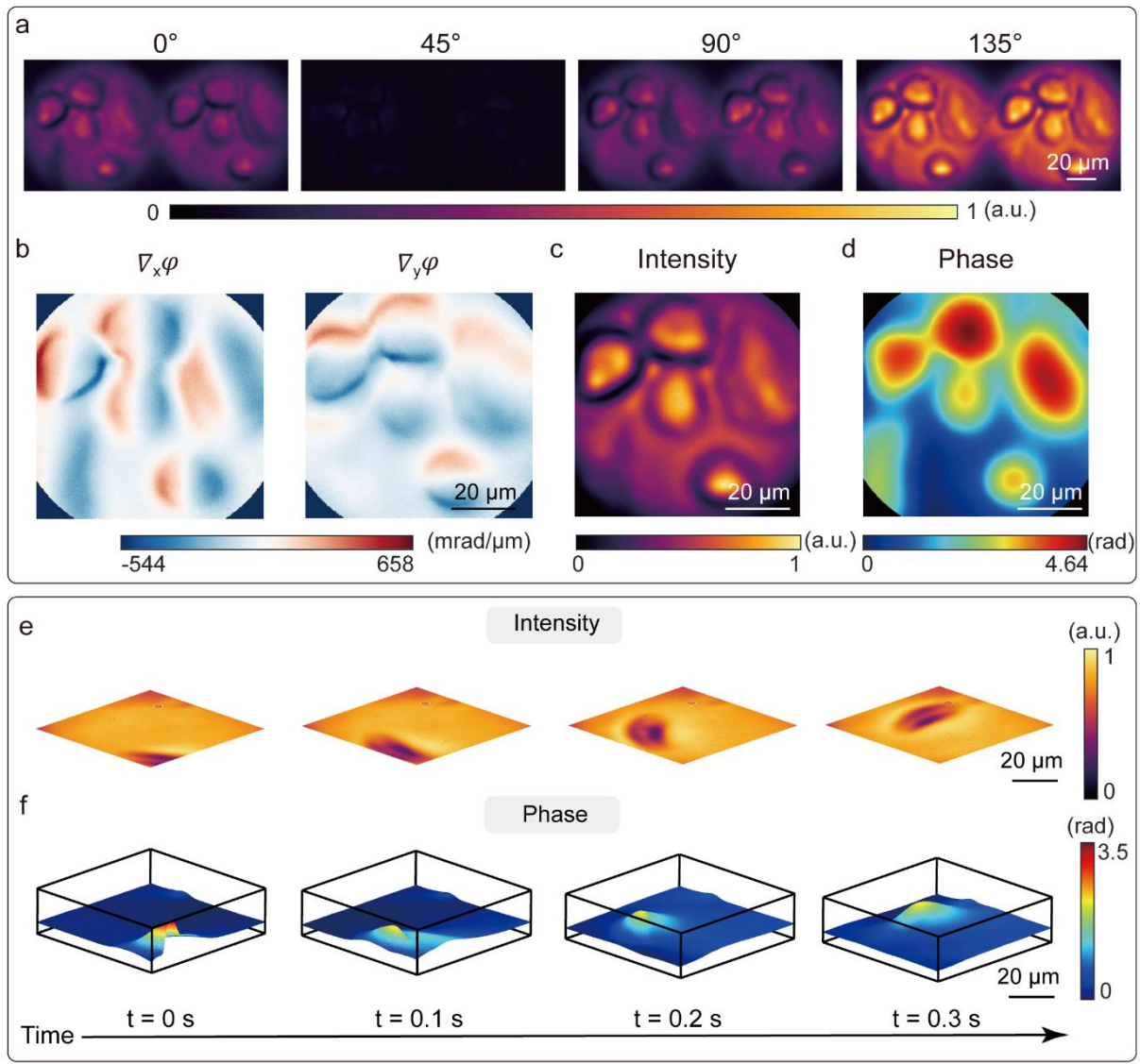

**Figure 5 | Complex amplitude microscopy for living cells. a**, Captured shearing interference images of living cancer cell MDA-MB-468 with the polarization channel along 0°, 45°, 90°, and 135°, respectively. **b**, Calculated phase gradients along the *x*- and *y*-direction, respectively. **c**, Reconstructed intensity image of the living cancer cells from the sum of 0° and 90° polarized light intensity. **d**, Reconstructed phase distribution of the living cancer cells. **e-f,** Selected frames of the measured intensity (**e**) and phase (**f**) distribution of the moving paramecium captured at a video frame rate of 10 Hz.

The complex amplitude imaging microscope can be used for the label-free observation of living cells. As an example, we imaged living cancer cell MDA-MB-468, as shown in Fig. 5a. with the reconstructed phase gradients and complex amplitude image shown in Fig. 5b-c, respectively. Since the complex amplitude image can be obtained in a single shot, the system is particularly suitable for observing the dynamic evolution of living cells and other transparent materials. Here, we also measured complex amplitude images of moving paramecium, with the video recorded at a frame rate of 10 Hz, as illustrated in Fig. 5f-g and Supplementary Video 1. In addition, we recorded the change in the surface morphology of an uncured UV adhesive sample subjected to an air cluster at a video frame rate of 50 Hz (Supplementary Video 2). The frame rate is limited only by the frame acquisition time of the polarization camera. Each frame, with a resolution of up to 281 × 281 pixels, can be reconstructed from raw measurements in approximately 0.41 seconds using a computer equipped with an AMD Ryzen 7 3700X CPU and 32GB RAM.

**Discussion**

In conclusion, this work combined the unmatched spatial- and polarization-multiplexing capability of a single-layer metalens with a commercially available polarization camera to allow deterministic reconstruction of complex amplitude light field in a single shot. The metalens can be designed as a stand-alone lens or to be coupled with a microscope system for a variety of imaging tasks. The system's minimum form factor and high robustness make it highly suitable for space-constrained imaging scenarios such as in endoscopes[18,52] and for portable, point-of-care diagnostics of fast-moving targets[53]. We also envision that the proposed system's capability to simultaneously record the complex amplitude light field information may pave the way for the construction of aberration-corrected 3D cameras[54]. Finally, by combining multiple compact complex amplitude imaging systems, one may build a long-sought, robust synthetic aperture lens at optical frequencies, similar to a synthetic aperture radar, for single-shot far-field super-resolution imaging[55].

**References**

1 M. Born & E. Wolf. *Principles of optics: electromagnetic theory of propagation, interference and diffraction of light*. 7th Edition, (Cambridge University Press, 2013).

2 G. Popescu. *Quantitative phase imaging of cells and tissues*. 1st Edition, (McGraw-Hill Education, 2011).

3 Y. Park, C. Depeursinge & G. Popescu. Quantitative phase imaging in biomedicine. *Nat. Photonics* 12, 578-589 (2018).

4 C. Zuo, J. Qian, S. Feng, W. Yin, Y. Li, P. Fan, J. Han, K. Qian & Q. Chen. Deep learning in optical metrology: a review. *Light Sci. Appl.* 11, 39 (2022).

5 Y. Gao & L. Cao. Iterative projection meets sparsity regularization: towards practical single-shot quantitative phase imaging with in-line holography. *Light Adv. Manuf.* 4, 37-53 (2023).

6 O. Guyon. Limits of adaptive optics for high-contrast imaging. *The Astrophysical Journal* 629, 592 (2005).

7 F. Zernike. How I discovered phase contrast. *Science* 121, 345-349 (1955).

8 W. Lang. *Nomarski differential interference-contrast microscopy*. (Carl Zeiss Oberkochen, 1982).

9 K. Creath. Phase measurement interferometry techniques. *Prog. Opt.* 26, 348-393 (1988).

10 V. H. Flores Munoz, N. I. Arellano, D. I. Serrano Garcia, A. Martinez Garcia, G. Rodriguez Zurita & L. Garcia Lechuga. Measurement of mean thickness of transparent samples using simultaneous phase shifting interferometry with four interferograms. *Appl. Opt.* 55, 4047-4051 (2016).

11 P. Marquet, B. Rappaz, P. J. Magistretti, E. Cuche, Y. Emery, T. Colomb & C. D. Depeursinge. Digital holographic microscopy: a noninvasive contrast imaging technique allowing quantitative visualization of living cells with subwavelength axial accuracy. *Opt. Lett.* 305, 468-470 (2005).

12 Z. Luo, J. Ma, P. Su & L. Cao. Digital holographic phase imaging based on phase iteratively enhanced compressive sensing. *Opt. Lett.* 44, 1395-1398 (2019).

13 G. Zheng, R. Horstmeyer & C. Yang. Wide-field, high-resolution Fourier ptychographic microscopy. *Nat. Photonics* 7, 739-745 (2013).

14 J. Sun, Q. Chen, J. Zhang, Y. Fan & C. Zuo. Single-shot quantitative phase microscopy based on color-multiplexed Fourier ptychography. *Opt. Lett.* 43, 3365-3368 (2018).

15  C. Zuo, J. Li, J. Sun, Y. Fan, J. Zhang, L. Lu, R. Zhang, B. Wang, L. Huang & Q. Chen. Transport of intensity equation: a tutorial. *Opt. Lasers Eng.* 135, 106187 (2020).

16  A. Greenbaum, W. Luo, T. W. Su, Z. Gorocs, L. Xue, S. O. Isikman, A. F. Coskun, O. Mudanyali & A. Ozcan. Imaging without lenses: achievements and remaining challenges of wide-field on-chip microscopy. *Nat. Methods* 9, 889-895 (2012).

17  C. Zuo, J. Sun, J. Zhang, Y. Hu & Q. Chen. Lensless phase microscopy and diffraction tomography with multi-angle and multi-wavelength illuminations using a LED matrix. *Opt. Express* 23, 14314-14328 (2015).

18  J. Sun, J. Wu, S. Wu, R. Goswami, S. Girardo, L. Cao, J. Guck, N. Koukourakis & J. W. Czarske. Quantitative phase imaging through an ultra-thin lensless fiber endoscope. *Light Sci. Appl.* 11, 204 (2022).

19  Y. Yang, W. Wang, P. Moitra, I. I. Kravchenko, D. P. Briggs & J. Valentine. Dielectric meta-reflectarray for broadband linear polarization conversion and optical vortex generation. *Nano Lett.* 14, 1394-1399 (2014).

20  M. Khorasaninejad, W. T. Chen, R. C. Devlin, J. Oh, A. Y. Zhu & F. Capasso. Metalenses at visible wavelengths: Diffraction-limited focusing and subwavelength resolution imaging. *Science* 352, 1190-1194 (2016).

21  W. T. Chen, A. Y. Zhu, V. Sanjeev, M. Khorasaninejad, Z. Shi, E. Lee & F. Capasso. A broadband achromatic metalens for focusing and imaging in the visible. *Nat. Nanotechnol.* 13, 220-226 (2018).

22  Y. Zhou, H. Zheng, I. I. Kravchenko & J. Valentine. Flat optics for image differentiation. *Nat. Photonics* 14, 316-323 (2020).

23  M. Liu, W. Zhu, P. Huo, L. Feng, M. Song, C. Zhang, L. Chen, H. J. Lezec, Y. Lu, A. Agrawal & T. Xu. Multifunctional metasurfaces enabled by simultaneous and independent control of phase and amplitude for orthogonal polarization states. *Light Sci. Appl.* 10, 107 (2021).

24  M. Pan, Y. Fu, M. Zheng, H. Chen, Y. Zang, H. Duan, Q. Li, M. Qiu & Y. Hu. Dielectric metalens for miniaturized imaging systems: progress and challenges. *Light Sci. Appl.* 11, 195 (2022).

25  X. Zhang, L. Huang, R. Zhao, H. Zhou, X. Li, G. Geng, J. Li, X. Li, Y. Wang & S. Zhang. Basis function approach for diffractive pattern generation with Dammann vortex metasurfaces. *Sci. Adv.* 8, eabp8073 (2022).

26  Y. Ni, C. Chen, S. Wen, X. Xue, L. Sun & Y. Yang. Computational spectropolarimetry with a tunable liquid crystal metasurface. *eLight* 2, 23 (2022).

27  Z. Shen, F. Zhao, C. Jin, S. Wang, L. Cao & Y. Yang. Monocular metasurface camera for passive single-shot 4D imaging. *Nat. Commun.* 14, 1035 (2023).

28  S. Wang, S. Wen, Z. L. Deng, X. Li & Y. Yang. Metasurface-Based solid poincare sphere polarizer. *Phys. Rev. Lett.* 130, 123801 (2023).

29  H. Kwon, E. Arbabi, S. M. Kamali, M. Faraji-Dana & A. Faraon. Computational complex optical field imaging using a designed metasurface diffuser. *Optica* 5, 924 (2018).

30  P. Huo, C. Zhang, W. Zhu, M. Liu, S. Zhang, S. Zhang, L. Chen, H. J. Lezec, A. Agrawal, Y. Lu & T. Xu. Photonic Spin-Multiplexing Metasurface for Switchable Spiral Phase Contrast Imaging. *Nano Lett.* 20, 2791-2798 (2020).

31  E. Engay, D. Huo, R. Malureanu, A. I. Bunea & A. Lavrinenko. Polarization-dependent all-dielectric metasurface for single-shot quantitative phase imaging. *Nano Lett.* 21, 3820-3826 (2021).

32  S. Yi, J. Xiang, M. Zhou, Z. Wu, L. Yang & Z. Yu. Angle-based wavefront sensing enabled by the near fields of flat optics. *Nat. Commun.* 12, 6002 (2021).

33  A. Ji, J.-H. Song, Q. Li, F. Xu, C.-T. Tsai, R. C. Tiberio, B. Cui, P. Lalanne, P. G. Kik, D. A. B. Miller & M. L. Brongersma. Quantitative phase contrast imaging with a nonlocal angle-selective metasurface. *Nat. Commun.* 13, 7848 (2022).

34  X. Wang, H. Wang, J. Wang, X. Liu, H. Hao, Y. S. Tan, Y. Zhang, H. Zhang, X. Ding, W. Zhao, Y. Wang, Z. Lu, J. Liu, J. K. W. Yang, J. Tan, H. Li, C. W. Qiu, G. Hu & X. Ding. Single-shot isotropic differential interference contrast microscopy. *Nat. Commun.* 14, 2063 (2023).

35  Y. Zhang, P. Lin, P. Huo, M. Liu, Y. Ren, S. Zhang, Q. Zhou, Y. Wang, Y. Q. Lu & T. Xu. Dielectric metasurface for synchronously spiral phase contrast and bright-field imaging. *Nano Lett.* 23, 2991-2997 (2023).

36  J. Liu, H. Wang, Y. Li, L. Tian & R. Paiella. Asymmetric metasurface photodetectors for single-shot quantitative phase imaging. *Nanophotonics* 12, 3519-3528 (2023).

37  T. Zhu, J. Huang & Z. Ruan. Optical phase mining by adjustable spatial differentiator. *Adv. Photonics* 2, 016001 (2020).

38  L. Wesemann, J. Rickett, J. Song, J. Lou, E. Hinde, T. J. Davis & A. Roberts. Nanophotonics enhanced coverslip for phase imaging in biology. *Light Sci. Appl.* 10, 98 (2021).

39  L. Li, W. Jia, C. Jin, S. Wang, Z. Shen & Y. Yang. Single-Shot Wavefront Sensing with Nonlocal Thin Film Optical Filters. *Laser Photonics Rev.* n/a, 2300426 (2023).

40  H. Kwon, E. Arbabi, S. M. Kamali, M. Faraji-Dana & A. Faraon. Single-shot quantitative phase gradient microscopy using a system of multifunctional metasurfaces. *Nat. Photonics* 14, 109-114 (2019).

41  J. Zhou, Q. Wu, J. Zhao, C. Posner, M. Lei, G. Chen, J. Zhang & Z. Liu. Fourier optical spin splitting microscopy. *Phys. Rev. Lett.* 129, 020801 (2022).

42  S. Liu, F. Fan, S. Chen, S. Wen & H. Luo. Computing liquid-crystal photonics platform enabled wavefront sensing. *Laser Photonics Rev.* 17, 2300044 (2023).

43  Q. Wu, J. Zhou, X. Chen, J. Zhao, M. Lei, G. Chen, Y.-H. Lo & Z. Liu. Single-shot quantitative amplitude and phase imaging based on a pair of all-dielectric metasurfaces. *Optica* 10, 619-625 (2023).

44  M. P. Kothiyal & C. Delisle. Shearing interferometer for phase shifting interferometry with polarization phase shifter. *Appl. Opt.* 24, 4439-4442 (1985).

45  K. Creath & G. Goldstein. Dynamic quantitative phase imaging for biological objects using a pixelated phase mask. *Biomed. Opt. Express* 3, 2866-2880 (2012).

46  L. Huang, M. Idir, C. Zuo, K. Kaznatcheev, L. Zhou & A. Asundi. Comparison of two-dimensional integration methods for shape reconstruction from gradient data. *Opt. Lasers Eng.* 64, 1-11 (2015).

47  J. Hardy & A. MacGovern. Shearing interferometry: a flexible technique for wavefront measurement. *Proc. SPIE* 0816, 180–195 (1987).

48  C. E. Shannon. Communication in the Presence of Noise. *Proceedings of the IRE* 37, 10-21 (1949).

49  K. Liu, J. Wang, H. Wang & Y. Li. Wavefront reconstruction for multi-lateral shearing interferometry using difference Zernike polynomials fitting. *Opt. Lasers Eng.* 106, 75-81 (2018).

50  T. Phan, D. Sell, E. W. Wang, S. Doshay, K. Edee, J. Yang & J. A. Fan. High-efficiency, large-area, topology-optimized metasurfaces. *Light Sci. Appl.* 8, 48 (2019).

51  P. Hariharan & D. Sen. Radial shearing interferometer. *Journal of Scientific Instruments* 38, 428 (1961).

52  H. Pahlevaninezhad, M. Khorasaninejad, Y. W. Huang, Z. Shi, L. P. Hariri, D. C. Adams, V. Ding, A. Zhu, C. W. Qiu, F. Capasso & M. J. Suter. Nano-optic endoscope for high-resolution optical coherence tomography in vivo. *Nat. Photonics* 12, 540-547 (2018).

53  Y. Ziv, L. D. Burns, E. D. Cocker, E. O. Hamel, K. K. Ghosh, L. J. Kitch, A. El Gamal & M. J. Schnitzer. Long-term dynamics of CA1 hippocampal place codes. *Nat. Neurosci.* 16, 264-266 (2013).

54	J. Wu, Y. Guo, C. Deng, A. Zhang, H. Qiao, Z. Lu, J. Xie, L. Fang & Q. Dai. An integrated imaging sensor for aberration-corrected 3D photography. *Nature* 612, 62-71 (2022).

55	K. Wang, Y. Q. Zhu, Q. C. An, X. C. Zhang, C. Peng, H. R. Meng & X. Y. Liu. Even sampling photonic-integrated interferometric array for synthetic aperture imaging. *Opt. Express* 30, 32119-32128 (2022).

## Methods

**Metalens fabrication.** The metalens was fabricated by a commercial service (Tianjin H-chip Technology Group). Fabrication began with a silicon-on-sapphire substrate, with a monocrystalline silicon film thickness of 600 nm. Electron beam lithography (JEOL-6300FS) was utilized to inscribe the metasurface pattern using a negative tone resist hydrogen silsesquioxane (HSQ). The pattern was then transferred to the silicon layer via reactive-ion etching, with HSQ directly serving as the mask. Finally, the HSQ resist was removed using buffered oxide etchant.

**PSF and diffraction efficiency measurement.** To measure the PSF of the fabricated metalens, we constructed a setup as shown in Supplementary Note 7. The illumination source consists of collimated light from a supercontinuum laser (YSL SC-PRO-7) and a bandpass filter (Thorlabs FB800-10) with a central wavelength of 800 nm and a bandwidth of 10 nm.

To estimate the diffraction efficiency of the metalens, the collimated laser beam was filtered by a linear polarizer (Thorlabs LPNIR100-MP2). An optical power meter (Thorlabs PM122D) with a pinhole of 2-mm diameter, equal to the metasurface diameter, was first placed in front of the metasurface to measure the power of the incident light $P_{inc}$. In the next step, to measure the power of the focused light $P_f$, a pinhole of 200-μm diameter was placed in front of the power meter. The position of the power meter was spatially scanned and maximized near the designed focal point of the metasurface. The diffraction efficiency of the metalens was estimated as $\eta = P_f / P_{inc}$.

**Complex amplitude imaging setup.** For the compact complex amplitude imaging system, the illumination source consists of a collimated LED, a narrowband filter with a center wavelength of 800 nm and a bandwidth of 10 nm, and a linear polarizer. The complex amplitude light field was captured by a commercial polarization camera (MV-CH050-10UP, Hikvision) through the metalens. The polarization camera is equipped with a polarization-sensitive CMOS sensor (IMX250LQR, SONY). Each sub-pixel of the sensor has a size of 3.45 μm, with 2 by 2 sub-pixels forming a macro-pixel for polarization detection along 0°, 45°, 90°, and 135°, respectively.

For the complex amplitude microscopy system, the collimated illumination source was generated via Köhler illumination. The light field at the image plane of the metalens was further magnified by a standard optical microscope consisting of an objective lens (Mitutoyo, MPLNAPO20XVIR 20×) and a tube lens with a focal length of 20 cm.

**Cell preparation.** Cancer cells MDA-MB-468 (Procell Life Science & Technology, China) were used in experiments. MDA-MB-468 cells were cultured using an incubator (Forma 381, Thermo Scientific, USA) at 37 °C in 5% CO2 and passaged three times a week with high-glucose Dulbecco's Modified Eagle's Medium (DMEM, Life Technologies, USA), supplemented with 10% fetal bovine serum (FBS, Life Technologies, USA) and 1% penicillin–streptomycin (Life Technologies, USA). After completing one cell passage, the cells were placed in the incubator and cultured for one day to obtain the sample of adherent cells for conducting the experiments.

## Data availability

All relevant data are available in the main text, in the Supporting Information, or from the authors.

**Acknowledgment**

This work was supported by the National Natural Science Foundation of China (62135008, 61975251), the China Postdoctoral Science Foundation (2023M731911), and the Guoqiang Institute, Tsinghua University. Y.Y. acknowledges Prof. Chao Zuo at Nanjing University of Science and Technology for helpful discussion.

**Author contributions**

L.L., S.W., and Y.Y. conceived this work. L.L., S.W and Y.Z. designed and characterized the metalens, built the complex amplitude imaging system, and performed the optical measurement; L.L., S.W., F.Z., Y.Z., S.W., H.C., Y.G., W.W., L.C., and Y.Y. analyzed the data; L.L., S.W., and Y.Y. wrote the manuscript. H.C. and W.W. prepared the cell samples. Y.Y. supervised the project.

**Competing interests**

The authors declare no competing interests.